\documentclass[twocolumn,amsmath,amssymb]{revtex4}

\usepackage[pdftex]{hyperref}
\hypersetup{bookmarksnumbered=true}
\hypersetup{bookmarksopen=true}
\hypersetup{bookmarksopenlevel=2}
\usepackage{doi}
\hypersetup{colorlinks=true}
\hypersetup{linkcolor=black}
\hypersetup{citecolor=black}
\hypersetup{filecolor=black}
\hypersetup{menucolor=black}
\hypersetup{urlcolor=black}

\usepackage{color}
\definecolor{darkgreen}{rgb}{0,0.5,0}
\definecolor{darkblue}{rgb}{0,0,0.5}
\definecolor{brown}{rgb}{0.98,0.92,0.73}
\definecolor{red}{rgb}{1,0,0}
\definecolor{yellow}{rgb}{1,1,0}
\definecolor{blue}{rgb}{0,0,1}
\definecolor{green}{rgb}{0,1,0}
\definecolor{purple}{rgb}{1,0,1}
\definecolor{gray}{rgb}{0.8,0.8,0.8}
\definecolor{black}{rgb}{0,0,0}
\definecolor{white}{rgb}{1,1,1}
\definecolor{gold}{rgb}{1.,0.84,0.}

\usepackage{graphicx}
\usepackage{caption}
\usepackage[]{siunitx}
\usepackage{mathptmx}
\usepackage{amsmath}
\usepackage{bm}
\usepackage{nicefrac}

\newenvironment{DIFnomarkup}{}{}

\def\narrowfigurewidth{0.6\columnwidth}
\def\figurewidth{0.8\columnwidth}

\newif\ifcmtr
\cmtrtrue
\ifcmtr 
\newcommand{\cmtr}[1]{ %
   [\color{red} \textbf{#1} \normalcolor]%
}%
\else
\newcommand{\cmtr}[1]{ %
}%
\fi

\ifcmtr 
\newcommand{\cmtrla}[1]{ %
   [\color{blue} \textbf{#1} \normalcolor]%
}%
\else
\newcommand{\cmtrla}[1]{ %
}%
\fi

\ifcmtr 
\newcommand{\cmtrth}[1]{ %
   [\color{purple} \textbf{#1} \normalcolor]%
}%
\else
\newcommand{\cmtrth}[1]{ %
}%
\fi

\ifcmtr 
\newcommand{\cmtrpl}[1]{ %
   [\color{green} \textbf{#1} \normalcolor]%
}%
\else
\newcommand{\cmtrpl}[1]{ %
}%
\fi

\begin{document}

\title{A thermodynamic description of turbulence\\ as a
  source of stochastic kinetic energy for 3D self-assembly}

\date{\today} 

\author{P.A. L\"othman$^{1,2*}$}
\author{T.A.G. Hageman$^{1,2*}$}
\author{M.C. Elwenspoek$^2$}
\author{G.J.M. Krijnen$^2$}
\author{M. Mastrangeli$^3$}
\author{A. Manz$^{1}$} 
\author{L. Abelmann$^{1,2}$}

\affiliation{
  $^1$ KIST Europe, Saarbr\"ucken, Germany\\
  $^2$ University of Twente, Enschede, The Netherlands\\
  $^3$ Delft University of Technology, Delft, The Netherlands\\
  $^*$ These authors contributed equallly to this work\\
}


\begin{abstract}

We investigate to what extent one can use a thermodynamic description
of turbulent flow as a source of stochastic kinetic energy for
three-dimensional self-assembly of magnetically interacting macroscopic
particles. We confirm that the speed of the objects in the flow field
generated in our system obeys the  Maxwell--Boltzmann distribution, and their
random walk can be defined by a diffusion coefficient following from
the Einstein relation. However, we discovered that the analogy with
Brownian dynamics breaks down when considering the directional
components of the velocity. For the vectorial components, neither the
equipartition theorem, nor the Einstein relation is obeyed. Moreover,
the kinetic energy estimated from the random walk of individual
objects is one order of magnitude higher than the value estimated from
Boltzmann statistics on the interaction between two spheres with
embedded magnets. These results show that introducing stochastic
kinetic energy into a self-assembly process by means of turbulent flow
can to a great extent be described by standard thermodynamic theory,
but anisotropies and the specific nature of the
interactions need to be taken into account.


\end{abstract}

\maketitle 
\section{Introduction}
Self-assembly is the autonomous organization of objects into a
structure without human intervention~\cite{Whitesides2002}. The final
properties of the assembled structure are exclusively based on the inherent
properties of the individual
objects~\cite{Elwenspoek2010}. Self-assembly has been achieved at
different scales and in multiple ways, including evaporation
induced self-assembly~\cite{Brinker1999}, ionic self-assembly~\cite{Faul2003}, self-assembly based on molecular recognition~\cite{Shenton1999, Fouquey1990}, DNA
self-assembly~\cite{Wang2013b}, and colloidal crystal
self-assembly~\cite{Kitaev2003}. The self-assembly of carbon nanotubes,
graphene oxide~\cite{Shimoda2002, Shao2014} and photonic
crystals~\cite{Xia2001}, as well as aspects such as the influence of
microgravity~\cite{Dag1997}, surfaces~\cite{Whitelam2015a}, and
supra-molecular engineering~\cite{Ciesielski2010, Faul2003}, have also been
evaluated.

In all varieties of and approaches to self-assembly, three elements
are common and critical: the characteristics of the individual
objects, such as their mutual binding forces and geometrical shapes,
the environment which may promote assembly with templates or other
forms of guidance, and the disturbing forces which oppose the binding
forces and thus allow the objects to explore the associated energy
landscape and find a global minimum of the energy~\cite{Pelesko2007}.

In our earlier paper~\cite{Hageman2018}, we introduced a macroscale
self-assembly process using magnetic interaction between the
objects and a turbulent flow as a source of disturbing forces. We
discovered that the random walk of the objects in the turbulent flow
can be succesfully described by thermodynamic theory. In the present paper, we
focus on the disturbing forces, and investigate up to what point the
thermodynamic description is valid. This question is highly relevant
for the self-assembly of objects that are so big that thermal energy
is no longer sufficient to drive the system into an energy minimum.

In self-assembly with very small objects, such as atoms or molecules,
the thermal energy $kT$ is adequate because according to the
equipartition theorem it corresponds to significant random (Brownian)
motion ($kT = \frac{1}{3} m \langle v^2\rangle$). However, for larger objects,
roughly above \SI{1}{\um}, thermal energy can no longer
provide a sufficient disturbing force~\cite{Elwenspoek2010}: the objects
would disintegrate rather than self-assemble.  Hence, to self assemble
macro-objects, alternative ways to provide disturbing forces are used,
mostly by some form of shaking~\cite{Mastrangeli2009}.

Even though shaking in the self-assembly of micro-meter sized objects has
the same function as the thermal energy in the self-assembly of atoms or
molecules, there are very distinct differences. In the first place,
shaking is a dissipative process. When we stop shaking, the random
motion of the objects comes to a halt. The energy we provide to the
system is partially transferred to the objects, but at the cost of
losing energy into heat along the way.

Secondly, shaking introduces a directionality into the disturbing
forces.  It is impossible to shake in all directions
simultaneously. One relies on random processes, such as collisions, to
randomize the direction of the forces. Since shaking is a dissipative
process, the effect of randomization has a limited
lifetime. Therefore, a signature of the initial direction of shaking
will always be present.

Both deviations from thermal energy, i.e. dissipation and directional
dependence, are present in all experiments where shaking is involved.
Our macroscale setup provides an easy way to study the nature of the
disturbing forces. The outcome is of general importance for the study
and implementation of self-assembly.

\subsection{Methods to provide disturbing forces}
Shaking the support on which the objects are placed is a common
method to provide disturbing forces in a self-assembly system.
Friction is used to transmit the energy to the object, as first shown
by Penrose~\cite{Penrose1957}.  The adhesion force between the objects and
the support varies with location, which causes a random motion.

This is used in two dimensions by placing objects on a moving platform
with a linear or orbital shaker~\cite{Tricard2013, Gruenwald2016,
  Ipparthi2018}, but also pseudorandom shaking has been
applied~\cite{Cademartiri2012}.  Rather than using stiction and
friction, one can place objects on smooth vibrating surfaces.  This is
exploited in experiments where the objects are placed at a liquid--air
interface, which is disturbed by shaking the
container~\cite{Haghighat2016, Miyashita2009, Jacobs2002,
  Bowden2001,Tien1997} or by ultrasound~\cite{Ahmed2014}. Random
motion is caused by collision with the wall of the container and between
the objects themselves.

In three dimensions, objects are shaken~\cite{Olson2015, Hacohen2015}
or rotated~\cite{Hosokawa1994} within containers; or one can submerge
the objects and agitate the fluid to induce particle motion, for
instance by
rotating~\cite{Terfort1997,Gracias2000,Clark2001,Shetye2010,Ilievski2011}
or shaking~\cite{Shetye2010,Onoe2007} the container, or by moving the
liquid itself by ultrasonic
agitation~\cite{Love2003,Shetye2010,Mastrangeli2014,Ilievski2011a,Woldering2016}
or a pulsating flow~\cite{Zheng2005,Goldowsky2013}.  An interesting
variation on this approach is to use diamagnetic levitation and
exploit the inertia of the objects to drive them out of energy
minima~\cite{Tkachenko2015}.

Instead of applying mechanical forces, one can apply disturbing energy
by an external magnetic field to magnetic
objects~\cite{Hosokawa1996,Grzybowski2002}. The magnetic field is
relatively uniform around each particle, therefore random motion
relies on collisions in this case. However, when the objects are
self-propelling, for instance by the decomposition of hydrogen
peroxide~\cite{Ismagilov2002}, random motion naturally occurs.

In the examples above, the focus of the research was on the products
of the self-assembly process, not on the disturbing force itself. It
is tempting to compare the energy provided to the system by some way
of shaking to thermal energy. For example, one can analyse the
distribution of the products generated by the self-assembly process by
Boltzmann statistics. Attempts have been made to introduce an ``effective
temperature''~\cite{Ilievski2011,Tricard2013} or ``propulsion
energy''~\cite{Ahmed2014}. In the present paper, we study in more detail the
nature of the disturbing forces, and to what extent they can be
described by standard thermodynamic theory.

\subsection{Self-assembly as a tool for technology and research}
We are interested in three-dimensional self-assembly as a
manufacturing method for next generation
electronics~\cite{Elwenspoek2010} and novel materials and devices
~\cite{Whitesides2005, Whitesides2002,Chidambaram2014}, as well as in
the advantages that studying self-assembly at the macroscale might
bring compared to the microscale.

Scanning electron microscopy and other microscopy images often show
the result of micro- or nanoscopic self-assembly, but they do not
reveal the process and dynamics behind it. We cannot see how the
structure formed or which pathways were taken. \textit{In situ}
transmission electron microscopy reveals self-assembly events rather
than the entire self-assembly process itself and its underlying
dynamics~\cite{Liu2013,Nakamura2017}. Moreover, it is impossible to
avoid the influence of the electron beam on the physical and chemical
properties of the particles as well as on the resolution of the
footage. It is likely that a change in the properties of the particle can lead to
altered particle interactions and therewith an altered
self-assembly. The effects of the electron beam are even more complex when
using liquid and gas environments at elevated temperatures rather than
vacuum~\cite{Taheri2016}.
 
Self-assembly often includes rapid phenomena, such as protein folding
or supramolecular or nanoparticle self-assembly, which can hardly be
observed directly at the microscopic level. Via macroscopic
experimentation we can overcome these obstacles by using
representative particles and an analogous macroscopic self-assembly
reactor, such as the one used in the research presented here. The dynamics of
the self-assembly process appears much slower at the macroscopic level
and can therefore be readily observed.

\subsection{Macroscale 3D self-assembly}
One aspect common to many three-dimensional self-assembly systems is
that the micro- or macroscopic particles stand or sediment onto the
bottom of the vessel when there is no agitation. Self-assembly studies
as a function of the strength of the disturbing forces are consequently
complicated. We therefore designed a system where the function of
levitating the objects is separated to a large extent from the
function of disturbing the system~\cite{Hageman2018}.

Levitation is achieved by introducing a flow opposite to the
force of gravity, counterbalancing the drop velocity of the objects
in the fluid. Agitation is achieved by intentionally introducing
a turbulent flow into the system by an asymmetric fluid inflow.

\begin{figure}
  \begin{center}
    \includegraphics[width=\narrowfigurewidth]{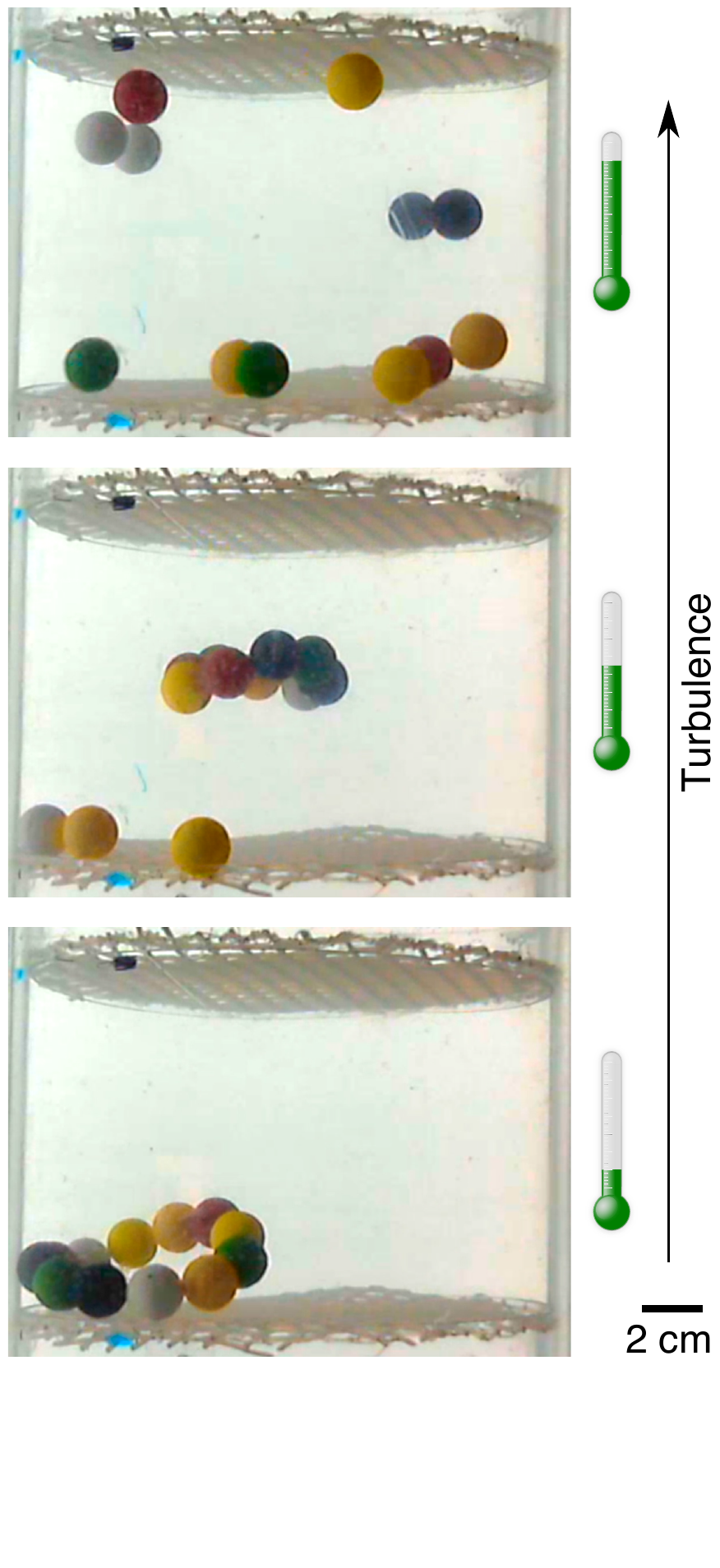}
    \caption{The effect of the degree of turbulence on the structure
      formation of 12 magnetic polymer spheres that self assemble
      in a vertical turbulent flow. Decreasing turbulence leads
      to increased structure formation (lines and rings of different
      lengths and shapes). At maximum turbulence only single spheres
      appear, whereas at minimum turbulence the lowest energy
      structure (a 12-sphere ring) appears. The local magnetic forces
      of each individual sphere interact as the spheres explore the
      energy landscape in order to find the configuration of lowest energy.
      A \href{https://youtu.be/x_p9X328W6Q}{video} is
      available in the Supporting Information. (The image of the
      thermometer to the right of each image was
      \href{https://commons.wikimedia.org/wiki/File:Green-Thermometer.svg}{adapted}
      with permission from the version by
      \href{https://commons.wikimedia.org/wiki/User:ARTunchained}{ARTunchained})}
    \label{fig:demo_multispheres}
  \end{center}
\end{figure}

As an example, Figure~\ref{fig:demo_multispheres} shows excerpts of a
video recording (available in the Supporting Information) of the
self-assembly of 12 polymer spheres of diameter \SI{2}{cm} with
embedded permanent magnets. The formation of the structure depends on the
degree of turbulence: at maximum turbulence, the spheres are
disconnected and start to form structures as the turbulence decreases. At
low turbulence, the minimum energy structure (ring) is formed. This is
a macroscopic representation of a microscopic quenching or cooling
sequence and nicely demonstrates the paths of self-assembly.

In this system, we thoroughly analysed the motion of the individual
objects~\cite{Hageman2018}. We have concluded that the random motion of the
objects can indeed to a large extent be described by standard
thermodynamic theory.

The speed of the objects obeys the Maxwell--Boltzmann distribution,
and there is a well defined diffusion constant, so that the Einstein
relation $kT=fD$ is obeyed.  Hence the concept of an effective
temperature has merit. However, we also found that there are
limitations to the validity of the thermodynamic description. For
instance, the speed distribution and the diffusion coefficient were
not dependent on the mass of the objects. Moreover, the kinetic energy
determined from the random motion was considerably higher than the
energy determined from the interaction between magnetic objects using
Boltzmann statistics. We suspect that the dissipative nature of
turbulence lies at the origin of this discrepancy. In this paper, we
present a thermodynamic description of self-assembly processes that
use turbulence as source of disturbing forces. We particularly
investigate the effect of the strength of the turbulence and analyse
the directional components of the random motion of the objects.


\section{Theory}
The analysis of particle trajectories and two-particle interactions
was introduced in~\cite{Hageman2018}. Here, we also analyse the
diffusion coefficient and velocity distribution for the projection of
the particle movement on the vertical axis ($z$), i.e. along the main
direction of the flow, and in the horizontal plane perpendicular to
the flow ($x,y$).  The diffusion of a particle in a confined space was
described in~\cite{Hageman2018} for one-dimensional movement along a
line segment. If the motions of the particle along the three projections are
uncorrelated, we can apply the same expression for the average squared
displacement:

\begin{equation}
	\label{eq:av_sq_disp_x}
	\left<x^2\right>  = \sigma_x^2 \left(1-\frac{x_\text{t} n(x_\text{t}, 
\sigma_x)}{N(x_\text{t}, \sigma_x)-\frac{1}{2}} \right),
\end{equation}

\noindent where $n(x, \sigma_x)$ is the normal distribution and $N(x, \sigma_x)$
is the cumulative normal distribution. For $x$, we can substitute the
$y$ or $z$ coordinate. the standard deviation of the
displacement is denoted by $\sigma_x$ and the variance $\sigma_x^2$ can be related to the
diffusion coefficient along a coordinate in one direction by

\begin{equation}
  \label{eq:Dx}
  \sigma^2_x=2D_xt.
\end{equation}

In~\cite{Hageman2018}, we used the Maxwell--Boltzmann distribution to
describe the distribution of the speed  in term of the most probable
speed (or mode) $v_\text{p}$. The distribution of the 
individual components of the velocity is Gaussian with the standard deviation
$\phi$ and zero mean
velocity component:

\begin{equation}
  \label{eq:vxdistr}
  p(v_x) =\frac{1}{\sqrt{2\pi \phi_x^2}} e^{-\frac{v_x^2}{2\phi_x^2}}
\end{equation}

\noindent where again $x$ can be substituted with $y$ or $z$.

In thermodynamic theory, the disturbing forces are described in terms
of the thermal energy $kT$. In the case of shaking, one can describe the
disturbing forces in terms of an `effective'
temperature~\cite{Ilievski2011,Tricard2013}, which is much higher
than the real temperature. However, the term `effective temperature'
implies that all aspects of the random motion of the components can be
described by standard thermodynamic theory. Therefore, we introduced
the term `disturbing energy' in our earlier
work~\cite{Hageman2018}. As with thermal energy, this `disturbing
energy' is in fact the stochastic contribution to the kinetic energy
of the particles. We believe it is more accurate to use the term
`stochastic kinetic energy', or simply `kinetic energy' when the
context is clear. These terms will be used throughout this paper.


\section{Methods}

The self-assembly reactor was introduced
in~\cite{Hageman2018}. The system has four inlet ports on the bottom
of the cylinder. For the present study, the inlet ports are equipped with
valves. This allows us to inject the water flow asymmetrically and
increase the turbulence. Two-way PVC ball valves (Type S6 DN40-14,
\SI{50}{mm} diameter, Praher Plastics Austria GmbH) were used.

Schematic front- and top-views of the reactor are shown in
Figure~\ref{fig:figure 2}. The valves can be opened between
\SI{0}{\degree} (fully closed) and \SI{90}{\degree} (fully
open). The maximum turbulence can be achieved by opening only one valve 
(right bottom image in Figure~\ref{fig:figure 2}, the three remaining
valves being opened by \SI{0}{\degree}) and the minimum turbulence by
opening all valves fully (left bottom image in Figure~\ref{fig:figure
  2}, \SI{90}{\degree}). For simplicity, between these two extremes we
decided to adjust the remaining three valves identically. A picture of
the self-assembly reactor and the valves is shown in the Supplementary
Material.

\begin{DIFnomarkup}
\begin{figure}
  \begin{center}
    \includegraphics[width=\figurewidth]
    {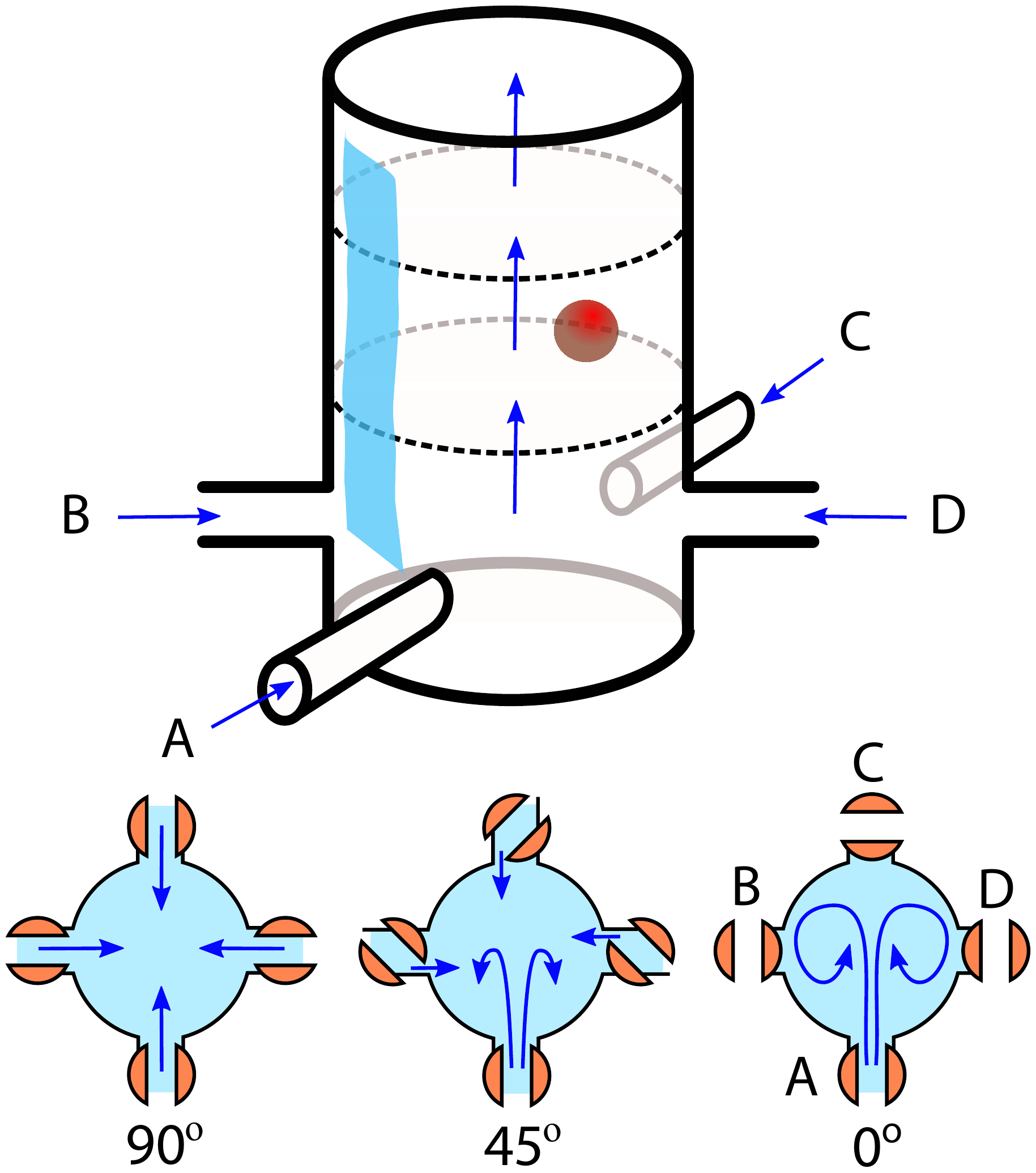}
    \caption{Top: Schematic view of the self-assembly reactor. An
      upward flow of water is inserted through 4 water inlets
      (A, B, C and D). The upward flow levitates a particle (red) and
      provides turbulence. The dotted circles in the middle of the
      reactor indicate the positions of the nets that are used as
      placeholders for the particle(s). The turbulence in the water
      flow is adjusted by closing valves B, C and D in the
      inlets. Bottom: Three valve settings are shown. For minimum
      turbulence all valves are open (left, \SI{90}{\degree} valve
      opening), for maximum turbulence valves B, C and D are fully
      closed (right, \SI{0}{\degree} valve-opening). A photograph of
      the setup and the valves is shown in the Supporting
      Information.}
    \label{fig:figure 2}
  \end{center}
\end{figure}
\end{DIFnomarkup}

To illustrate the effect of an asymmetric inflow, we inserted an air
diffusor at the bottom of the reactor and used it to generate a
curtain of bubbles. Since the bubbles tend to follow the flow,
they can be used to image the flow pattern. Figure~\ref{fig:bubbles}
shows two frames of a high speed movie of the bubble flow. The movie
(available in the Supplemental Material) was recorded at
\SI{240}{fps}, and is set to play back \num{10} times slower. The flow
speed of the water was \SI{10.5(5)}{cm/s}.

The movie demonstrates that there is always turbulence in the
reactor. This is in agreement with the speed of the flow and the
diameter of the tube. We estimate the Reynolds number of the flow to
be in the order of \num{18e3}, which is indeed substantially above the
laminar flow regime. When the flow is fully asymmetric, the turbulence
increases and swirls are visible in the flow. At the experimental flow
rate it takes only around \SI{0.5}{s} for the water front to travel
the \SI{18}{cm} from the bottom to the top of the reactor.  This
observation accounts for that turbulence and the vortice size
distribution seem not to vary with increasing reactor height.

\begin{DIFnomarkup}
\begin{figure}
  \begin{center}
    \includegraphics[width=\figurewidth]
    {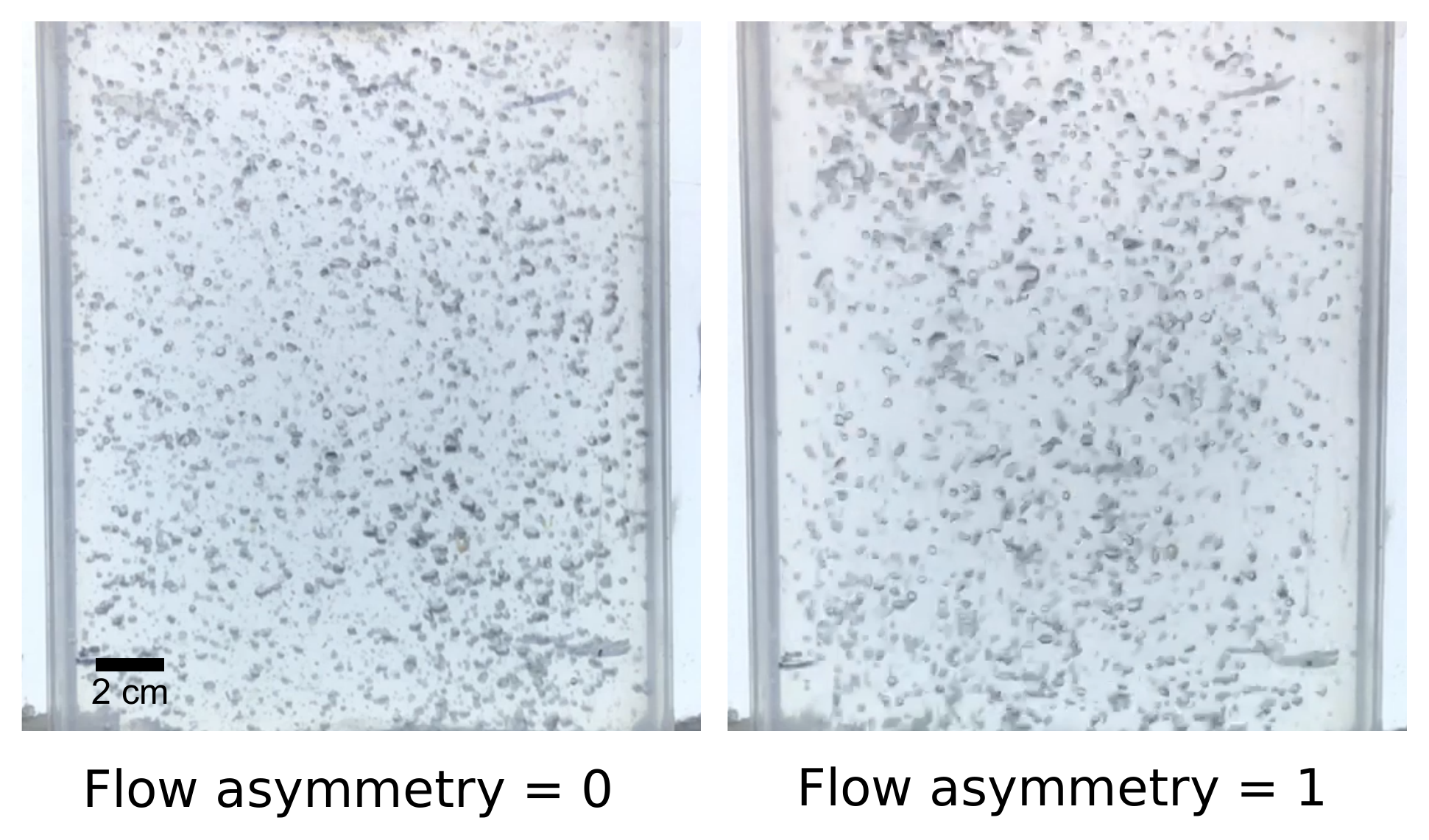}
    \caption{Bubble curtain in the upward water flow for all valves
      fully open (left) and three out of four closed (right). In the
      right image, an increase in turbulence can be observed, which is
      very apparent in the movie that is available as Supplemental
      Material.}
    \label{fig:bubbles}
  \end{center}
\end{figure}
\end{DIFnomarkup}

\subsection{Flow calibration}
\begin{figure}
  \begin{center}
    \includegraphics[width=\figurewidth]{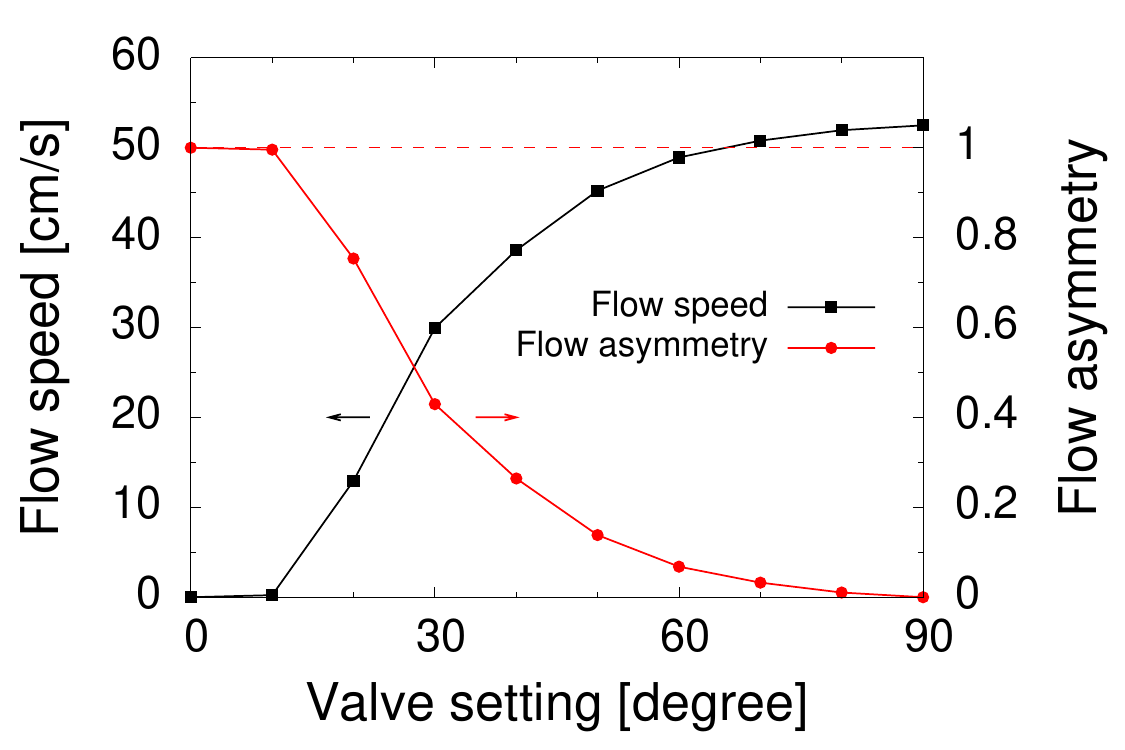}
    \caption{Flow speed through the reactor at maximum pump effort as
      a function of the valve angle of one valve, when the other
      three remain closed.}
    \label{fig:valve_calibration}
  \end{center}
\end{figure}

We expect the turbulence in the cylinder to be proportional to the
asymmetry in the inflow. The inflow is determined by the angular
position of the four valves. To analyze the relation between valve
settting and flow, we measured the flow through the cylinder as a
function of opening angle $\theta$ of one single valve, see
Figure~\ref{fig:valve_calibration}. In this measurement, the other
three valves are closed and we used the maximum pump effort. We
accordingly define a dimensionless measure for the asymmetry of the
flow:

\begin{equation}
	\mathrm{flow\ asymmetry} = 1- \frac{f(\theta)}{f(90^\circ)},
\end{equation}

\noindent where $f(\theta)$ is the speed of the flow of the water through the valves
controlled during the experiments, i.e. at opening angle $\theta$. This
way, at minimum turbulence, when all valves are fully open, the flow
asymmetry is defined to be \num{0}, whereas at maximum turbulence, when three
valves are closed, the flow asymmetry is defined to be \num{1}.

\subsection {Particles}

The particles were identical, as in~\cite{Hageman2018}, being
\SI{18.80(7)}{mm} diameter ($2R$) polymer (ABS) spheres with a
\num{3.80}$\times$\SI{3.80(5)}{mm} cylindrical NdFeB permanent magnet
placed in the centre of each sphere. The mass of the particles was
\SI{4.14(1)}{g}. In order to calculate the stochastic kinetic energy,
we estimate the effective mass $m^*$ to be \SI{5.9(1)}{g} by adding
\SI{50}{\percent} of the mass of the displaced water~\cite{Hageman2018}.

The terminal drop velocity $v_\text{t}$ of the particles, measured in
a column of water without flow, was \SI{37(1)}{cm/s}, from which we
estimate the drag coefficient $C_d$ to be \SI{0.35(6)}{}, which is in
agreement with the theoretical value of
\num{0.39}~\cite{Brown2003}. In order to estimate the stochastic
kinetic energy of the particles, we introduced an effective drag
coefficient~\cite{Hageman2018}

\begin{equation}
  \label{eq:1}
  f=\rho_\text{fluid}C_\text{d}\pi R^2 v_\text{t}
\end{equation}

\noindent where $\rho_\text{fluid}$ is the density of water
(\SI{998(3)}{kg/m^3}). At this flow velocity, the value of the
effective drag coefficient is \SI{35(4)}{g/s}.

In the experiment, we set the flow velocity of the water to
\SI{18.6(3)}{m/s}. The flow velocity was lower than the terminal drop
velocity of the sphere in order to avoid having the spheres touch the
top net and get trapped there. Since during the measurement of the position,
the sphere is levitating in front of the camera, we employ the
terminal drop velocity, rather than the flow velocity, for the
estimate of the effective drag coefficient.

\subsection {Reconstruction}
Two synchronized cameras were used for video recordings, as described
in~\cite{Hageman2018}. The particles were observed under different degrees
of turbulence. For each setting, videos were recorded for
\SI{15}{\minute} for single sphere experiments and \SI{30}{\minute}
for two sphere experiments. Both the 3D trajectories of a single
sphere and the distance between two spheres were reconstructed via
custom written Matlab scripts.

\subsection{Measurement precision}
To determine the diffusion coefficient, the trajectory of the particle
in the turbulent flow was observed, as described in~\cite{Hageman2018}.
Each trajectory longer than \SI{0.5}{\second} was fitted to the
diffusion model described in~\cite{Hageman2018}. These values were
averaged for a large number of trajectories to obtain an estimate of
the diffusion coefficient. The precision of the estimate increases
with the number of measurements, which is expressed in the standard
error (the standard deviation of the fit divided by the square of the
number of fits).

\begin{figure}
  \begin{center}
    \includegraphics[width=\figurewidth]{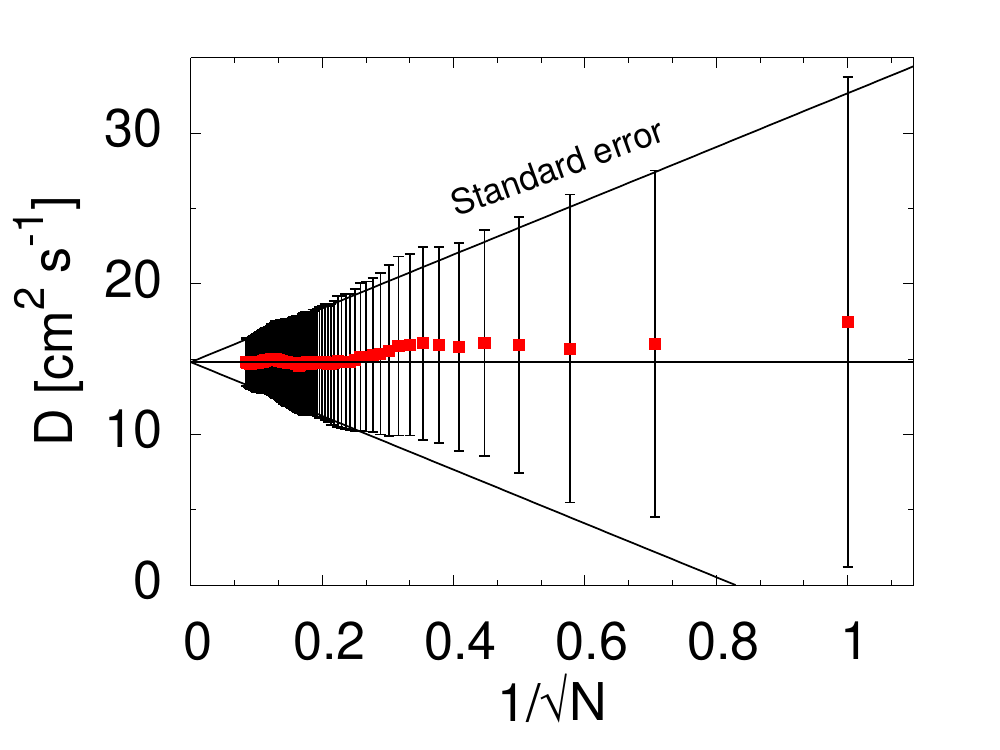}
    \caption{The estimated diffusion coefficient (red dots) and
      1$\sigma$ confidence interval as a function of the inverse
      square root of the number of trajectories ($N$). }
    \label{fig:errors}
  \end{center}
\end{figure}

To validate this process, we determined the diffusion coefficient for
sets of data with varying numbers of trajectories.  The result is shown
in Figure~\ref{fig:errors}, where the error bars represent the
estimate of the diffusion coefficient as a function of the number of
trajectories $N$. As expected, the estimated value converges (to about
\SI{15}{\square\centi\meter\per\second}) with increasing number of
measurements. The error bars indicate the 1$\sigma$ confidence limit
on the estimate (we are \SI{68}{\percent} confident that the diffusion
coefficient lies between the bars).  Figure~\ref{fig:errors} shows
that indeed the precision of the estimate increases with the square
root of the number of trajectories.

From this measurement, we conclude that for a 1$\sigma$ confidence
limit of \SI{5}{\percent} of the estimated value, we need at least
\num{570} trajectories.  We obtain approximately \SI{80}{trajectories}
of \SI{0.5}{s} duration per minute. The total measurement time per
experiment should therefore be at least \SI{7}{\minute}. To be on the
safe side, the duration of the experiments in this study was increased to
\SI{15}{\minute}.

Even though the estimate of the overall diffusion coefficient is well
behaved, we observe a large scatter on the estimates of the diffusion
coefficient of the individual components, see
figure~\ref{fig:diffusion}. This scatter cannot be explained by the
uncertainty on the estimates themselves. We conclude therefore that for the
individual components, there must be other sources of uncertainty. For
linear fits to the individual components (figures~\ref{fig:diffusion}
and ~\ref{fig:kTvel} bottom), we therefore ignore the error estimates in the
linear fits.

\section{Results}
\label {sec: Results}

We observed the movement of a single sphere and the interaction
between two spheres in the reactor, and determined the
stochastic kinetic energy as a function of the flow asymmetry, applying the
methods introduced in~\cite{Hageman2018}.  We also
investigated the directional dependency in the velocity distribution.

\subsection {Relation between flow asymmetry and stochastic kinetic energy}
\label{sec:ResultskT}

We observed the influence of turbulence on the kinetic behaviour of a
single particle in terms of the most probable speed $v_\text{p}$ and
its diffusion coefficient, as well as the interaction between two
particles. From these observations, we determined the relation between
the flow asymmetry and the stochastic kinetic energy.

\subsubsection{Influence of flow asymmetry on speed}
\begin{figure}
  \begin{center}
    \includegraphics[width=\figurewidth]{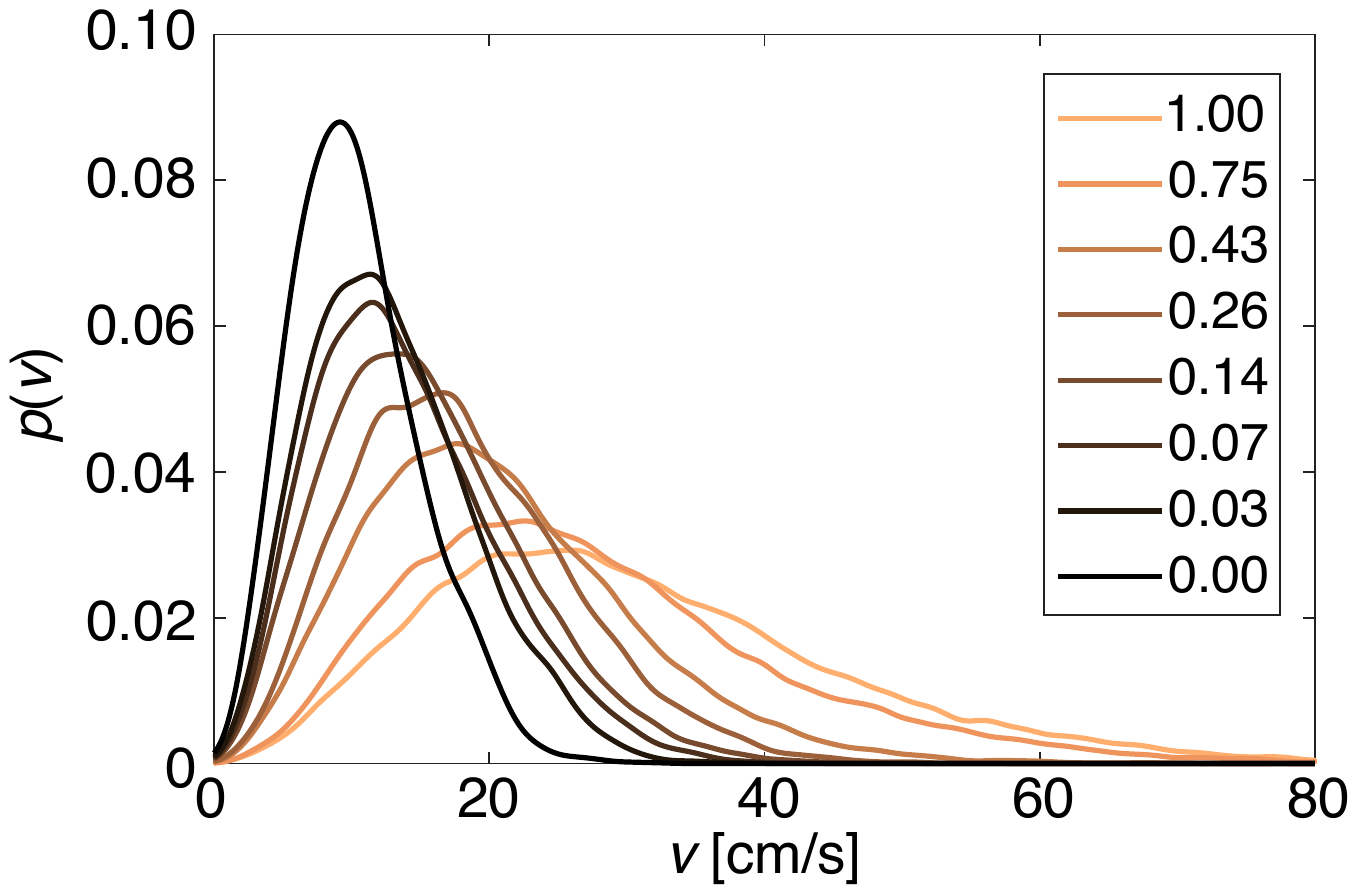}
    \caption{The distributions in the speed of a single particle for different
      settings of the flow asymmetry in the reactor show a
      Maxwell--Boltzmann-like distribution. The distribution was
      obtained via a kernel density estimation using a Gaussian kernel
      with $\sigma=1$ \si{\centi\meter\per\second}. Increased
      turbulence leads to a higher average speed.}
    \label{fig:velocity_dist}
  \end{center}
\end{figure}

Figure~\ref{fig:velocity_dist} shows the distribution of the speed of a
particle in a turbulent flow for various settings of the flow
asymmetry. These probability density functions were obtained by a
kernel density estimation using a Gaussian kernel with a standard
deviation of \SI{1}{\centi\meter\per\second}. With increasing flow
asymmetry there is an increase in the speed of the particle.

\begin{figure}
  \begin{center}
    \includegraphics[width=\figurewidth]{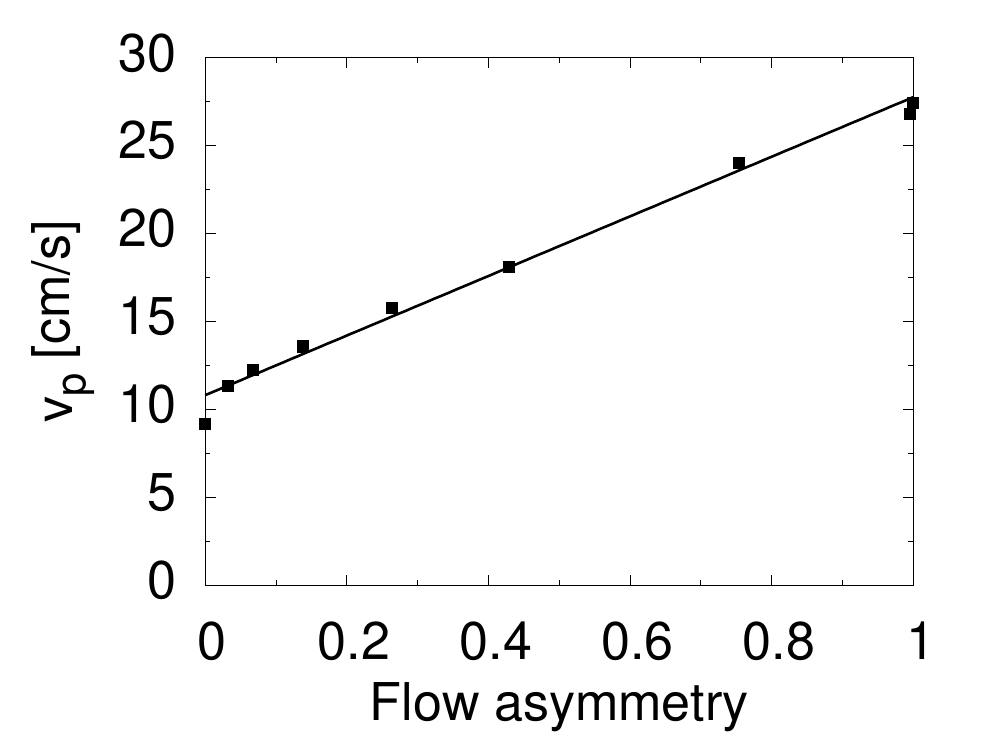}
    \includegraphics[width=\figurewidth]{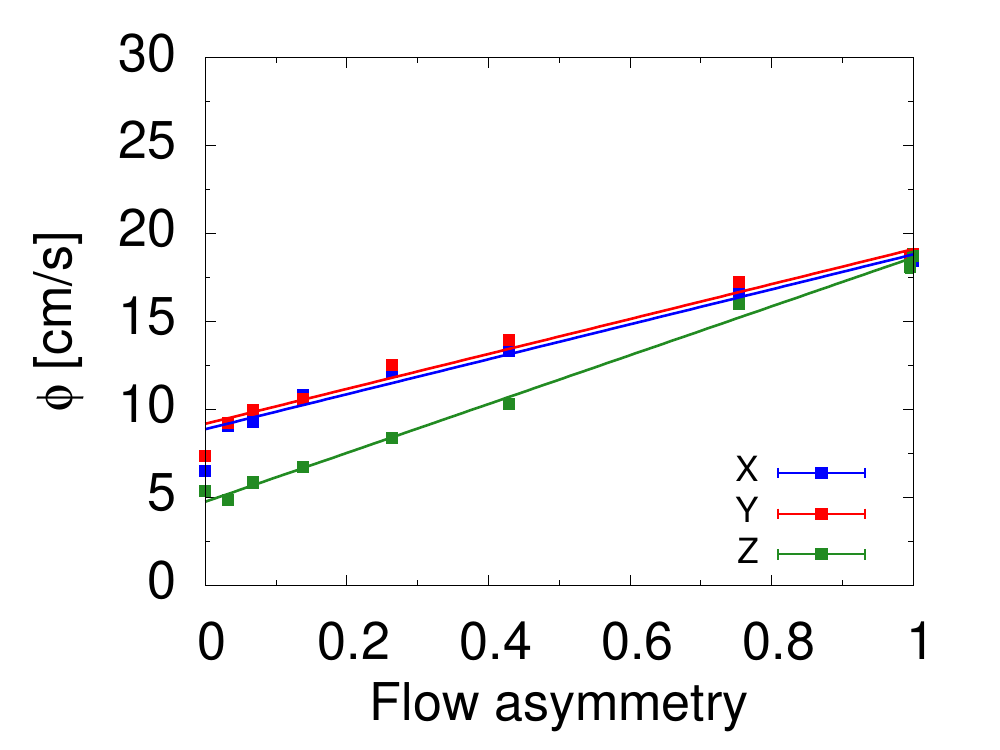}
    \caption{\textbf{Top:} the mode $v_\text{p}$ of the
      Maxwell--Boltzmann distribution of the speed of the particle as
      a function of the asymmetry of the flow. The relation is
      approximately linear. The most probable speed (mode) increases
      by almost a factor of three, indicating that the turbulence is
      increased. \textbf{Bottom:} the standard deviation of the
      horizontal ($x$, $y$) and vertical ($z$) components of the
      particle velocity as a function of the asymmetry of the flow.
      The width of the velocity distribution in the vertical direction
      is significantly smaller than in the horizontal direction for a
      flow asymmetry below \num{0.5}.}
   \label{fig:velocity}
  \end{center}
\end{figure}

The measured distribution of the speed was fitted to a
Maxwell--Boltzmann distribution, as described in~\cite{Hageman2018},
with only the most probable speed $v_\text{p}$ as the fitting
parameter. Figure \ref{fig:velocity} $v_\text{p}$ shows as a function
of the flow asymmetry. This relation is approximately linear. Over the
full range of the available flow asymmetry, the speed varies by a
factor of three from approximately
\SIrange{10}{30}{\centi\meter\per\second}.

\subsubsection{Influence of the flow asymmetry on the diffusion coefficient}

\begin{figure}
  \begin{center}
   \includegraphics[width=\figurewidth]{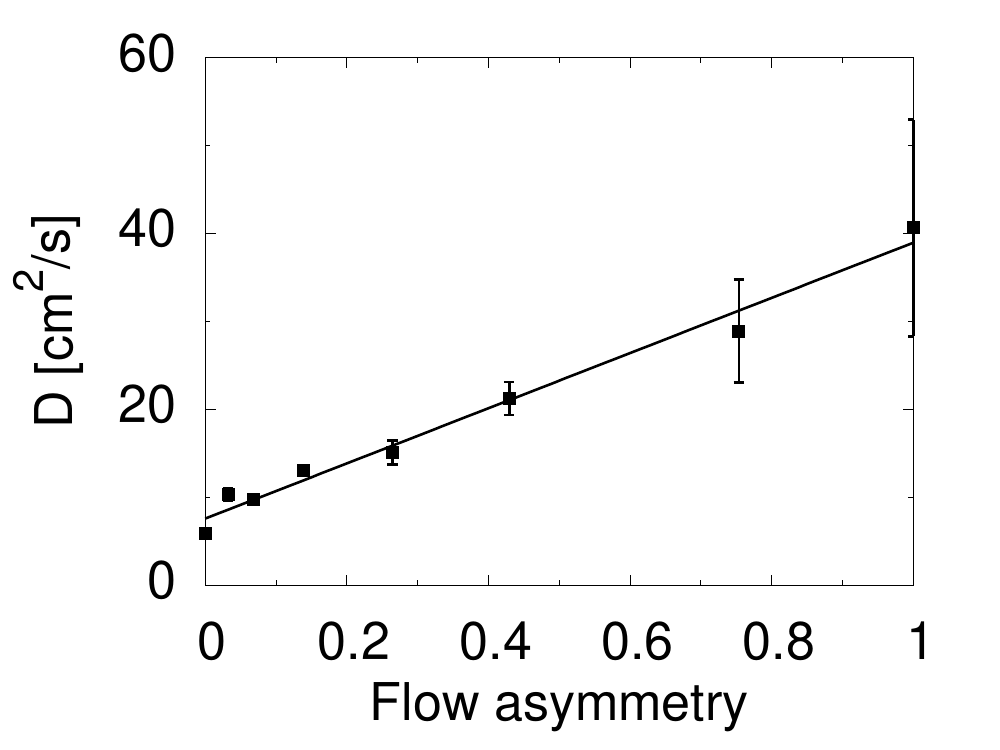}
    \includegraphics[width=\figurewidth]{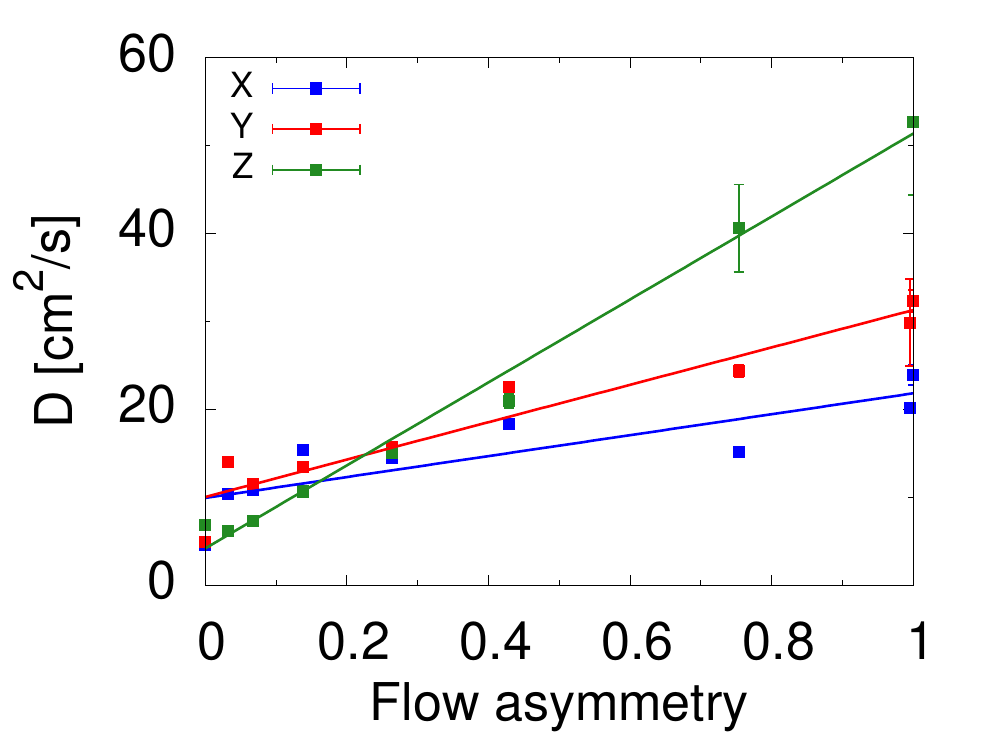}  
    \caption{\textbf{Top:} the diffusion coefficient of the motion of
      a single particle as a function of the asymmetry of the flow. The relation
      is roughly linear. \textbf{Bottom:} the diffusion coefficient
      for each direction as a function of the asymmetry of the flow. Above an
      asymmetry of \num{0.5}, the difference between the components is
      fairly large, but reduces significantly for lower
      turbulence.}
      \label{fig:diffusion}
\end{center}
\end{figure}

The diffusion coefficient was estimated by fitting a confined random
walk model to the measured average squared displacement
~\cite{Hageman2018}. The latter was obtained by averaging the squared
displacements of the trajectories with a duration of
\SI{2}{\second}. Figure \ref{fig:diffusion} shows the diffusion
coefficient as a function of the asymmetry of the flow. As in the case of
the speed, the diffusion increases roughly linearly with the asymmetry of the flow,
now by a factor of 6 from approximately
\SIrange{7}{44}{\square\centi\meter\per\second} (minimum and maximum
turbulence respectively).

\subsubsection{Influence of the asymmetry of the flow on the stochastic kinetic energy}
As described in~\cite{Hageman2018}, the speed distribution as well
as the diffusion coefficient of a single sphere can be related to the
stochastic kinetic energy through $kT=\nicefrac{1}{2}m^*v_\text{p}^2$
and the Einstein relation $kT=fD$.

A third method for obtaining the stochastic kinetic energy makes use of
the interaction between the particles.  When two particles are inserted in the reactor,
they connect and disconnect intermittently. The ratio between the time
they are connected and disconnected depends on their magnetic
interaction energy and the kinetic energy in the
system. In~\cite{Hageman2018} a method is described to determine this
energy from the distribution of the observed particle distances. This
method is more precise and fundamentally more correct than the method
based on the durations of the connections and disconnections.

\begin{figure}
  \begin{center}
    \includegraphics[width=\figurewidth]{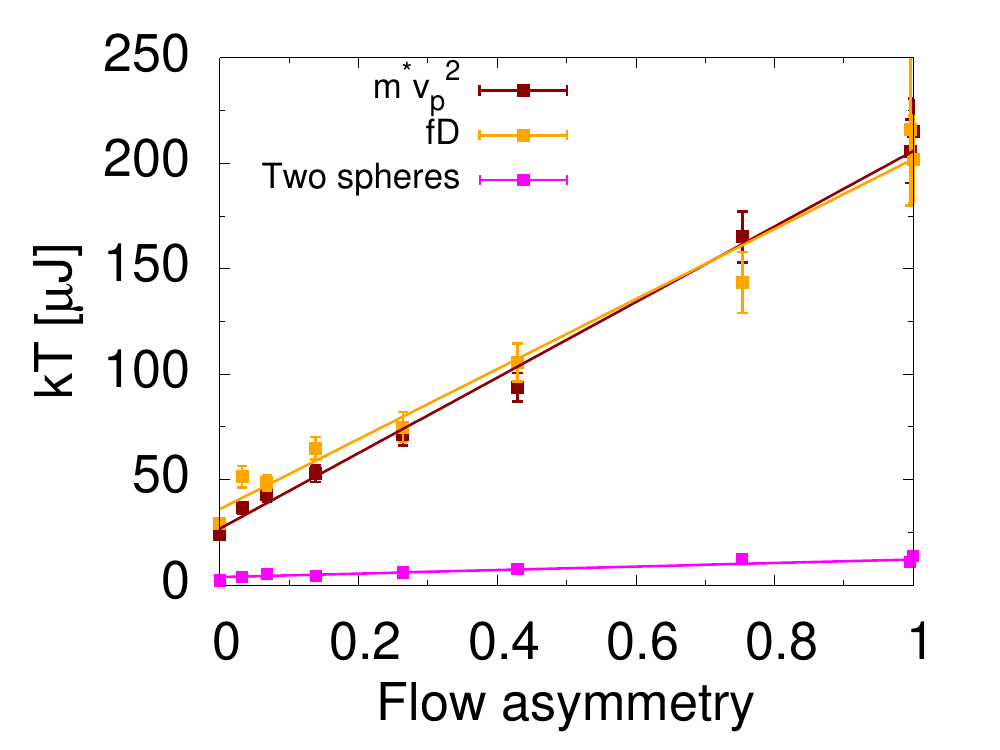}
    \includegraphics[width=\figurewidth]{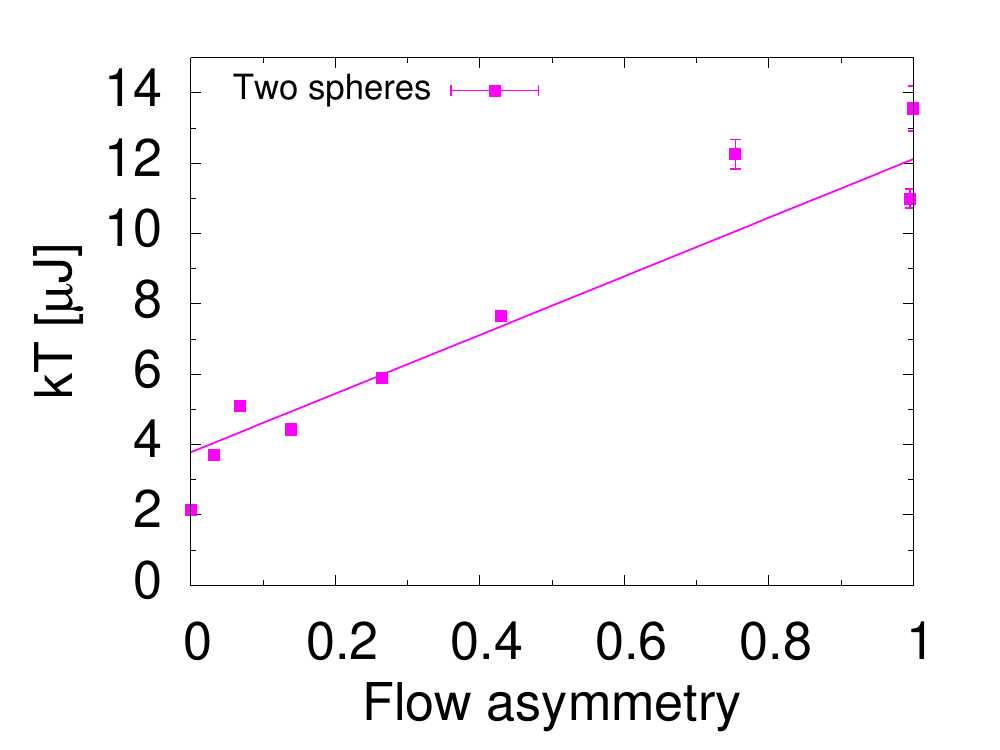}
    \caption{Stochastic kinetic energy ($kT$) as determined from the
      single-sphere experiments, using the diffusion coefficient
      (Einstein relation $fD$), using $\nicefrac{1}{2}m^*v_\text{p}^2$,
      and using the interaction between two spheres. The kinetic energy
      determined from the speed distribution agrees very well with
      that obtained from the diffusion coefficient. However, the values
      obtained from the single-sphere experiments are an
      order of magnitude higher than those from the two-sphere experiment
      (enlarged at the bottom).}
    \label{fig:kT}
  \end{center}
\end{figure}

Figure~\ref{fig:kT} shows all estimates for the stochastic kinetic
energy, plotted together. The relation between $kT$ and the asymmetry  of the flow
fits well to a linear function in all three cases. The 
coefficients of fit are listed in Table~\ref{tab:fits}. The estimates of the
kinetic energy from the single sphere experiments are very
similar. However, like in~\cite{Hageman2018}, these values are an
order of magnitude higher than the values obtained from the two-sphere
experiments. For both the single- and the two-sphere experiments, the kinetic
energy increases with increasing flow asymmetry. The increase is
higher by a factor of approximately two for the single-sphere experiment (for
the single-sphere experiment, the increase $a/b$= \SI{5(2)}{}, and for
the two-sphere one, $a/b$=\SI{2.1(7)}{}).

\begin{table}
  \centering
  \begin{ruledtabular}
  \begin{tabular}[t]{ccc}
    & slope ($a$) & offset ($b$)\\
    $m^*v_p^2$ &  \num{179(9)} & \num{27(2)} \\
    $fD$ &  \num{166(9)} & \num{36(5)} \\
    Boltzmann & \num{3.8(3)} & \num{8(2)}\\
  \end{tabular}
  \end{ruledtabular}
  \caption{Results of linear fits ($ax+b$) of the stochastic kinetic
    energy to the flow asymmetry, using the kinetic energy $m^*v_p^2$,
    diffusion coefficient $fD$, and interaction between two spheres
    using Boltzmann statistics.}
  \label{tab:fits}
\end{table}

\begin{figure}
  \begin{center}
    \includegraphics[width=\figurewidth]{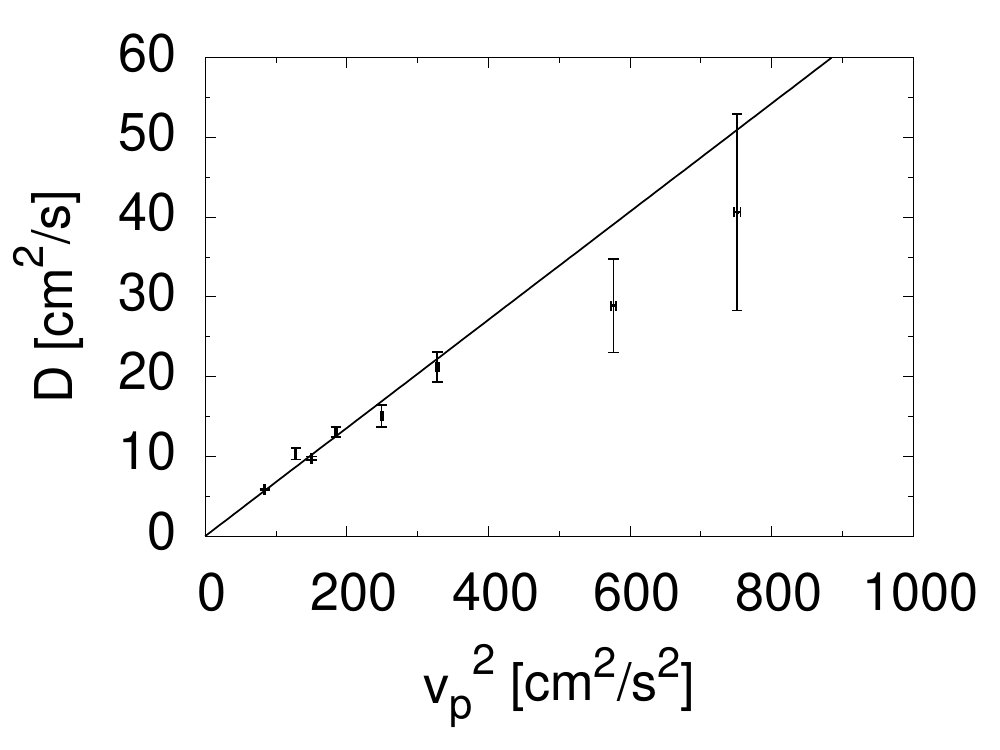}
    \caption{Diffusion coefficient versus the square of the mode of
      the speed distribution.  The ratio is constant, with a slope
      of \SI{68(1)}{ms}. This value is close to the theoretical
      prediction of $\frac{1}{2}\tau_\text{v}$=\SI{83(11)}{ms}.}
    \label{fig:D_vs_Vp2}
  \end{center}
\end{figure}

The flow asymmetry is a parameter specific to our setup. To generalize
the results, we analysed the ratio between the diffusion coefficient
and the speed. Equating the stochastic kinetic energy obtained from
the diffusion coefficient to that obtained using the velocity, we obtain
$D=\frac{1}{2}\tau_\text{v}v_\text{p}^2$, where $\tau_\text{v}=m^*/f$
is the characteristic time separating the ballistic regime from the Brownian
motion regime~\cite{Bian2016, Hageman2018}. Figure~\ref{fig:D_vs_Vp2}
shows that the plot of $D$ versus $v_\text{p}^2$ is
indeed approximately linear. The slope of the fit is \SI{68(1)}{ms}, which is
close to the value of $\frac{1}{2}\tau_\text{v}$=\SI{83(11)}{ms} and
very similar to the value \SI{0.1}{s} reported by Ilievski \emph{et
  al.} for \SI{1}{cm} blocks in a turbulent
flow~\cite{Ilievski2011}. This result is encouraging, considering that
in our analysis the effective mass $m^*$ and effective drag
coefficient $f$ are measured independently.

\subsection {Directional dependence of stochastic kinetic energy}
\label{sec:ResultsDirectional}

The water flow is directed from the bottom to the top in the reactor
in order to counteract the action of gravity on the particles. It is
therefore expected that the vertical ($z$) component of the motion of the particle
deviates from the horizontal ($x$ and $y$)
components. Additionally, there might be an asymmetry in the
$xy$-plane as well, since the flow is injected asymmetrically at high
turbulence. These effects are present both in the velocity
distribution as well as in the diffusion coefficient.

\subsubsection{Directional dependence of velocity}
\begin{figure}
  \begin{center}
    \includegraphics[width=\figurewidth]{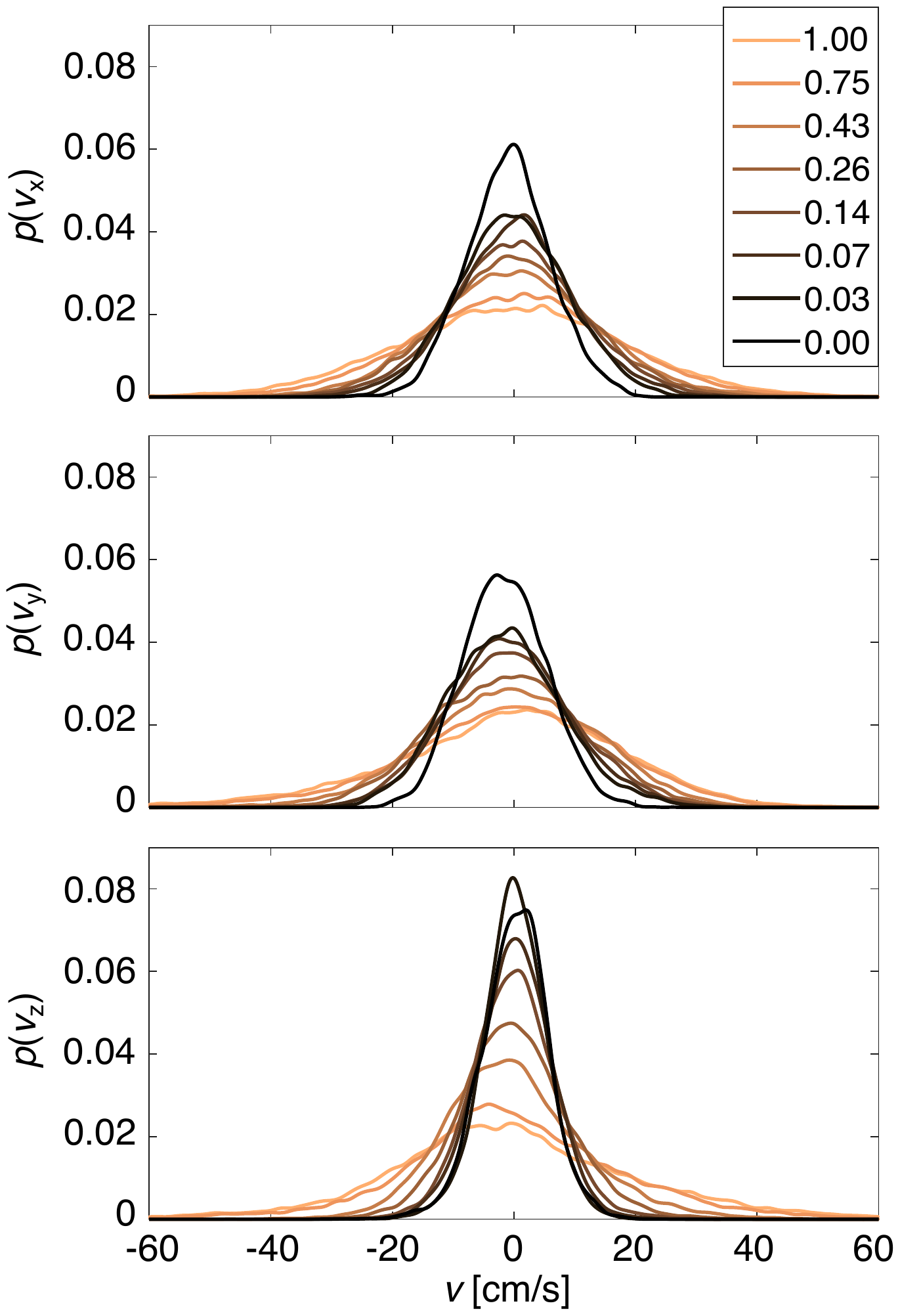}
    \caption{The $x$, $y$ and $z$ components of the velocity of a
      single particle for different settings of the flow asymmetry in the
      reactor show a Gaussian-like distribution. The distribution was
      obtained via a kernel density estimation using a Gaussian kernel
      with $\sigma=1$ \si{\centi\meter\per\second}. Increased
      turbulence leads to a wider velocity distribution. The $z$ component
      of the velocity has a significantly lower distribution width than the horizontal
      components.}
    \label{fig:velocity_dist_XYZ}
  \end{center}
\end{figure}

Figure~\ref{fig:velocity_dist_XYZ} shows the velocity distribution of
a particle in a turbulent flow for various settings of flow
asymmetry. Also in this case the graphs were obtained by a kernel
density estimation using a Gaussian kernel with a standard deviation
of \SI{1}{\centi\meter\per\second}. With increasing flow asymmetry,
the velocity distribution becomes wider, so that the average
absolute value of the velocity increases. The velocity in the horizontal directions
is similar, but the vertical velocity is significantly lower for settings of low
asymmetry.  In accordance with theory, a normal distribution
(Equation~\ref{eq:vxdistr}) was fitted to the measurements. The
standard deviation $\phi$ is plotted in
{Figure~\ref{fig:velocity}}. There is no significant difference in the
horizontal directions ($x$ and $y$ components), and there seems to be
no correlation of the difference with the degree of asymmetry of the flow. For flow
asymmetry, below \num{0.5}, the velocity in the $z$-direction is
significantly lower, by up to a factor of 2.

\subsubsection{Directional dependence of the diffusion coefficient}
Figure~\ref{fig:diffusion} shows the diffusion coefficients
along the three different directions. Even though the data is
scattered, the values for the horizontal directions only differ
moderately.  The diffusion coefficient in the $z$-direction, however,
shows a much stronger dependence on the flow asymmetry, diving below 
that for the horizontal components for low flow asymmetry, and vice versa.

\subsubsection{Directional dependence of stochastic kinetic energy}
\begin{figure}
	\begin{center}
          \includegraphics[width=\figurewidth]{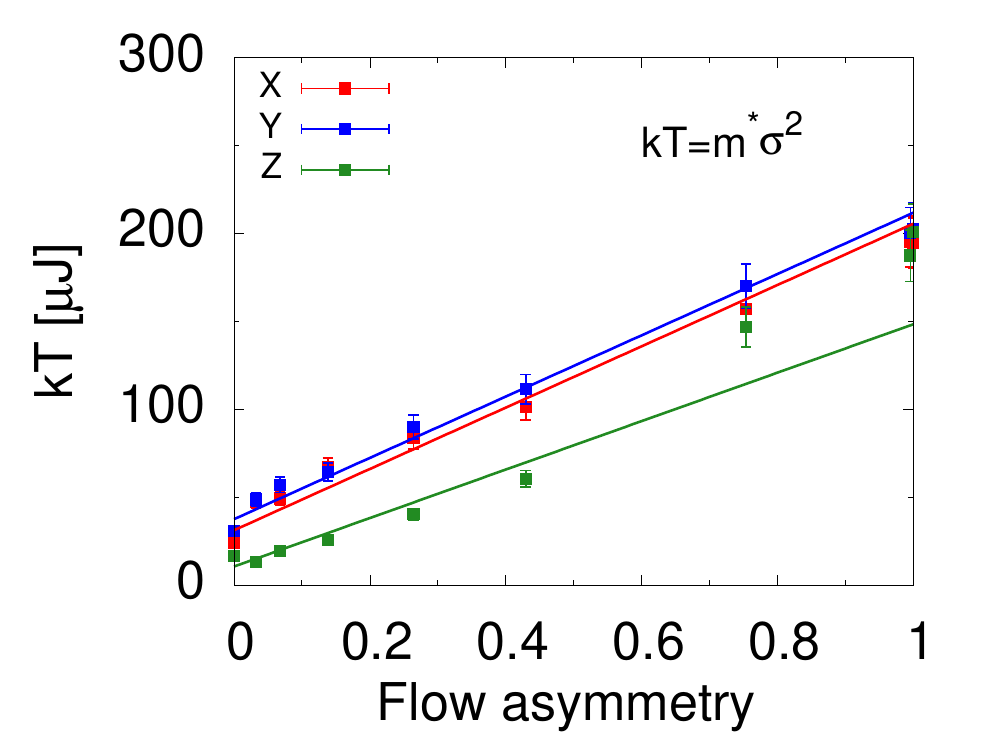}
          \includegraphics[width=\figurewidth]{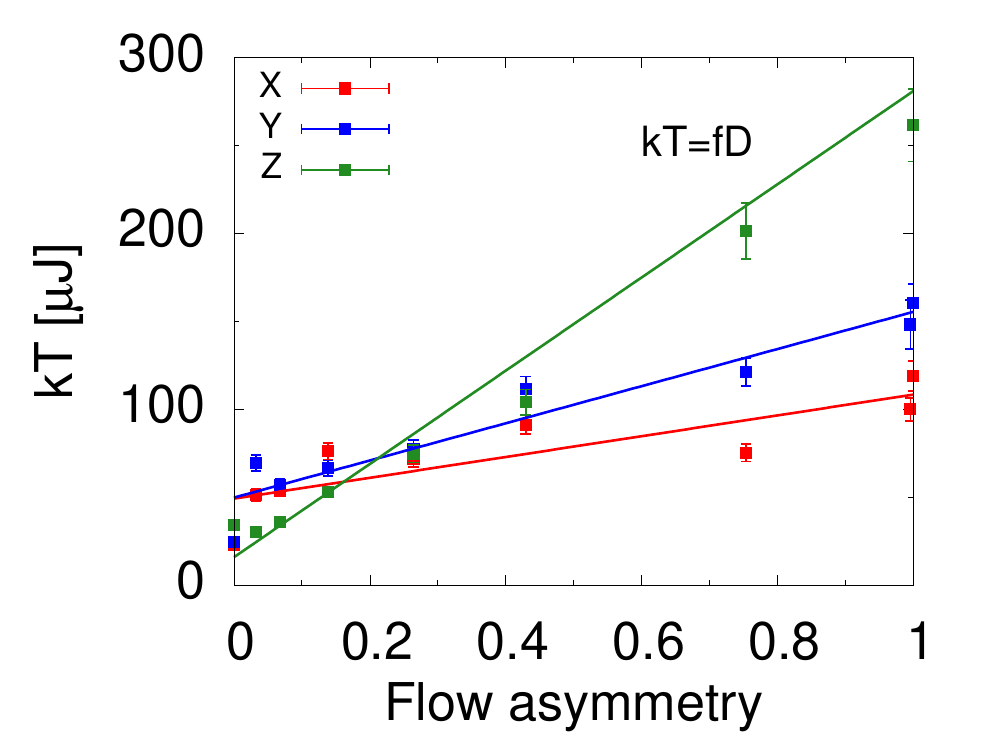}
          \caption{The stochastic kinetic energy in the three
            different directions, derived from the velocity (top) and
            from the diffusion coefficient (bottom). In general, the
            energy in all directions increases with increasing flow
            asymmetry. The estimate based on the diffusion coefficient
            suffers from scatter, even though the measurements
            themselves are precise (error bars). When considering the
            velocity distribution, the energy in the horizontal ($xy$)
            directions are equal, but twice as large as the
            energy in the vertical ($z$) direction for flow
            asymmetries below \num{0.5}. }
      \label{fig:kTvel}
  \end{center}
\end{figure}

The kinetic energy can be derived from the velocity and diffusion
coefficients for the individual $x$-, $y$- and $z$-components as
shown in Figure~\ref{fig:kTvel}. For clarity, two graphs are
plotted, one of the estimate based on the kinetic energy
($kT = m^* \phi^2$, top) and one for the estimate based on the
Einstein relation ($kT = fD$, bottom). Of course these graphs show
similar trends as {Figures~\ref{fig:velocity}} and
{~\ref{fig:diffusion}}, as the particle mass and drag coefficient do
not change between the measurements: the velocity and the diffusion
coefficient fully determine the shape of these curves.

Using the velocity distribution, the stochastic kinetic energy is
equal in the horizontal plane within the measurement error. However, for flow
asymmetries below \num{0.5}, the energy in the vertical direction is
less than one-half of that in the horizontal directions. When using the
diffusion coefficient to determine the directional dependence of the
stochastic energy, the scatter in the estimates is higher even though
the fits are still precise (error bars). The estimated values of the
stochastic kinetic energy are in the same range as that of the estimates
based on the velocity, especially at low flow asymmetry. The analysis
suggests that an estimate based on the velocity distribution is to be
preferred when considering the directional dependence, as it suffers
less from scatter.

\section{Discussion}
Our experiments clearly show that the particle velocity, diffusion coefficient, and stochastic kinetic energy increase with the degree of turbulence.  When three of the four inlet valves are gradually closed, the inflow becomes more asymmetric and the turbulence increases. This process adds to the turbulence created by the geometry of the reactor.

Creating inflow asymmetry is a practical way to change turbulence and mimic temperature changes at the micro- and nano-scale. The analogy between turbulent motion and thermal fluctuation is quite intriguing. There are, however, at least two aspects where the analogy between turbulence and thermal fluctuation does not hold: isotropy and spatial frequency power density.

\subsection{Anisotropy in turbulent flow}

The experiments show that one cannot ignore the directionality of the turbulent flow field. In analogy with temperature fluctuation, we would have to conclude that the temperature in the system is anisotropic.

Judging from the observations on the directional dependence, an increase in turbulence has a more pronounced influence on the vertical direction. The differences in the velocity, diffusion coefficient, and stochastic kinetic energy between the $x$- and $y$-directions are mild, especially compared to those of the $z$-component.  The latter also has a higher range between its minimum and maximum values at the extremes of the flow asymmetry.

The anisotropy of the stochastic kinetic energy is more pronounced when derived from the Einstein relation than it is when derived from the velocity. This might have to do with the nature of the velocity: the theory of diffusion assumes a purely random process. A bias might affect how this velocity contributes to the observed displacement over time, and in this way to the validity of Equation~\ref{eq:Dx}.

There is a region around the flow asymmetry minima of \num{0.5}, in which the directional dependence is a minimum. We are confident that the directional differences between the variables can be minimized by proper technical reconstruction of the self-assembly reactor. Altering the number and location of the inlet tubes and valves might be one possible option to create a more homogeneous three-dimensional flow field in which multi-particle self-assembly can be realized.

\subsection{Length scale}

The value of the stochastic kinetic energy determined via the two-particle experiments is an order of magnitude lower than that obtained from the diffusion or velocity of a single particle. We believe that this is because a greater part of the provided energy contributes rather to the motion of the single particles than to their close interaction. There is a vortex hierarchy in turbulent flow (Richardson cascade~\cite{Richardson1926}).  Hence the larger vortices in a turbulent flow must first break up into smaller vortices, until viscous forces become significant and dissipate energy.

Furthermore, we assume that the asymmetrical introduction of the turbulent flow causes a macroscopic swirl with a diameter similar to the diameter of the tank at the bottom of the cylinder. We think that the swirl moves upward in a screw-like manner and may actually represent the largest vortice in a Richardson cascade. We observed how a stream of air bubbles moved upwards in a screw-like manner with a screw-diameter approaching the diameter of the tank as they were introduced at the bottom of the self-assembly reactor.

Due to the Richardson cascade, there is an energy transfer from larger vortices to smaller ones. The energy is not uniformly distributed over the length scales, and drops off at shorter length scales~\cite{Hwang2012}.  So, in contrast to thermal fluctuation, the equipartition theorem does not hold for the energy spectrum (i.e. turbulence ``noise'' is not white). When we consider velocity or diffusion, we take into account all vortices, whereas for the two-sphere experiment, only vortices with length scales on the order of the sizes of the  particles contribute to their separation. So it is not surprising that the value of the stochastic kinetic energy derived from that experiment is lower.

\subsection{Implication for self-assembly}

\textcolor{red}{The observed deviations from standard thermodynamics may not specific to our experimental configuration, but may be present in all self-assembly experiments where some form of agitation other than  thermal energy is used.}

The directionality in stochastic kinetic energy may lead to an anisotropic
growth of the assemblies, which could even be desirable. On the other
hand, if the assemblies are free to rotate in the fluid, the growth
may be isotropic.

However, the fact that the stochastic kinetic energy decreases with
decreasing length scale can become an obstacle. If the assembly grows,
the disturbing forces increase as well. This will limit the maximum
size of the achievable assemblies. There are two measures one can
take. In the first place, one can gradually reduce the power of
the shaking over time, so that as the assembly increases in size, the
disturbing force generated by the shaking action remains the same.
Alternatively, one can ensure that when the assembly grows, the forces
that are required to break it increase as well. In one-dimensional
assemblies, such as the lines and rings in Figure~\ref{fig:demo_multispheres}, this is not the case. But in
two- and three-dimensional self-assembly, the binding forces indeed
increase with an increasing number of parts in the assembly.


\section{Conclusions}

We have analysed the movement of centimeter-sized spheres in a vertically biased turbulent flow field, and compared this movement with the thermodynamic theory for
Brownian motion.

We found that the speed of a single sphere in the turbulent flow
obeys the Maxwell--Boltzmann distribution and its movement can be described
by a confined random walk with a well defined diffusion coefficient,
identical to that of the Brownian motion of a sub-micron particle in a fluid.

We created an asymmetric inlet flow, which introduces an additional
turbulence on top of the turbulence resulting from the high velocity of the flow
and the geometry of the reactor. With increasing asymmetry, both the diffusion coefficient and the
mode of the distribution of the speed increase. In analogy to
the thermodynamic thermal energy term $kT$, we defined a stochastic
kinetic energy, using either the effective mass of the sphere and the
the mode of the speed distribution ($\tfrac{1}{2}m^*v_\text{p}^2$) or the drag coefficient
and diffusion coeffient ($fD$). These values equal within
the measurement error over the entire range of turbulence and increase
from \num{25} to \SI{200}{\micro J} with increasing asymmetry in the
inlet flow.

The analogy with Brownian motion breaks down when considering the
vectorial components of the velocity. The water flow is upwards, to
compensate for the drop velocity of the spheres. As a result, at low
turbulence, the vertical component of the velocity of the sphere is twice as
large as that in the lateral direction. This difference disappears at higher
turbulence. In contrast, the diffusion coefficient in the vertical
direction is approximately equal to that of the lateral directions at
low turbulence, but is higher by almost a factor of 3 at high turbulence. So
neither the equipartion theorem nor the Einstein relation are obeyed
when considering the individual components.

The analogy with standard thermodynamics also breaks down when
comparing the stochastic kinetic energies for different
experiments. We estimated the stochastic kinetic energy from the
interaction between two spheres with embedded magnets. This energy
again increases with increasing turbulence, but is an order of
magnitude lower than the value obtained from the single sphere
experiment (\num{2.1} up to \SI{13.6}{\micro J}). For self-assembly
studies, this value of the stochastic kinetic energy is more relevant.

These results show that the shaking due to a turbulent flow can, to a
certain extent, be described by standard thermodynamic theory, but directional dependencies should be taken into account, and one
cannot simply translate the value of the stochastic kinetic energy
from one expermental configuration to another. This result is of
importance for the self-assembly of objects with sizes above the
micrometer range, where thermal motion is no longer effective and some
form of shaking needs to be applied to drive the system into the
minimum energy state.


\section*{Acknowledgements}


The authors would like to thank Remco Sanders for building the setup and L\'eon Woldering for initial work on the project.
We further thank Michael Dirnberger for thoughtful insights, and
Nikodem Bienia and Gayoung Kim for their useful contribution to the
scientific work.


\bibliographystyle{apsrev_modified_doi}
\bibliography{../../../PaperBase/paperbase}

\end{document}